\documentclass[review]{elsarticle}

\usepackage{lineno,hyperref}
\usepackage{bm}
\usepackage{mathrsfs}
\usepackage{amssymb,amsmath}
\usepackage{graphicx}
\usepackage{color}
\modulolinenumbers[5]
\usepackage{geometry}
\journal{}

\begin{document}

\begin{frontmatter}

\title{Variable-mixing parameter quantized kernel robust mixed-norm algorithms for combating impulsive interference\tnoteref{mytitlenote}}
\tnotetext[mytitlenote]{* Corresponding author at: School of Electrical Engineering, Southwest Jiaotong University, Chengdu, Sichuan, 610031, PR China.\\E-mail\;addresses: $\mathrm {lulu@my.swjtu.edu.cn (L.\;Lu), hqzhao\_swjtu@126.com (H.\;Zhao), chenbd@mail.xjtu.edu.cn (B.\;Chen)}$.\\}

\author{Lu Lu$^{a}$, Haiquan Zhao$^{a*}$, Badong Chen$^{b}$}
\address{a)School of Electrical Engineering, Southwest Jiaotong University, Chengdu, China}
\address{b)School of Electronic and Information Engineering, Xi'an Jiaotong University, Xi'an,China.}

\begin{abstract}
Although the kernel robust mixed-norm (KRMN) algorithm outperforms the kernel least mean square (KLMS) algorithm in impulsive noise, it still has two major problems as follows: (1) The choice of the mixing parameter in the KRMN is crucial to obtain satisfactory performance. (2) The structure of the KRMN algorithm grows linearly as the iteration goes on, thus it has high computational complexity and memory requirements. To solve the parameter selection problem, two variable-mixing parameter KRMN (VPKRMN) algorithms are developed in this paper. Moreover, a sparsification algorithm, quantized VPKRMN (QVPKRMN) algorithm is introduced for nonlinear system identification with impulsive interferences. The energy conservation relation (ECR) and convergence property of the QVPKRMN algorithm are analyzed. Simulation results in the context of nonlinear system identification under impulsive interference demonstrate the superior performance of the proposed VPKRMN and QVPKRMN algorithms as compared with the existing algorithms.
\end{abstract}


\end{frontmatter}


\section{Introduction}

Kernel method has received increasing attention in machine learning and adaptive signal processing. The main idea of the kernel method is to transform the input data into a high-dimensional feature space via a reproducing kernel Hilbert space (RKHS). Some successful applications motivate to improve the robustness of the nonlinear adaptive filter (e.g., support vector machine (SVM) \cite{scholkopf2002learning,ishida2018development} and the kernel principal component analysis (KPCA) \cite{scholkopf1998nonlinear,fezai2018online}). 

Recently, the kernel adaptive filters became popular due to their modeling capabilities in the feature space. By using Mercer kernels \cite{liu2011kernel}, many linear filters have been recast in high-dimensional reproducing kernel Hilbert spaces (RKHSs) to yield more powerful nonlinear extensions, such as the kernel recursive least squares (KRLS) algorithm \cite{engel2004kernel,van2006sliding,liu2009extended,santos2017outlier}, the kernel least mean square algorithm (KLMS) \cite{liu2008kernel,takeuchi2016better,wang2017kernel} and the kernel affine projection algorithm (KAPA) \cite{liu20082kernel,albu2017low}. These algorithms have been successfully applied in nonlinear active noise cancellation \cite{liu2011kernel,liu2008kernel} and nonlinear acoustic echo cancellation (NLAEC) \cite{gil2013nonlinear}. 

Although the above-mentioned kernel adaptive filters achieve improved performance, they are not suitable for online applications, as their structures grow linearly with the number of processed patterns. In the past years, some sparsification techniques that constrain the growth of the network size were proposed \cite{liu2011kernel,engel2004kernel,richard2009online,chen2012quantized,xu2015quantised,wang2017quantized,flores2017set}. In 2012, the quantized KLMS (QKLMS) algorithm has been successfully applied to static function estimation and time series prediction \cite{chen2012quantized}. It has a mechanism to utilize the redundant input data, which is helpful to achieve a better accuracy and a more compact network with fewer centers.

On the other hand, in most signal processing applications, impulsive noise exists widely. It is well known that the impulsive noises have infinite variance, which makes the traditional $l_2$-norm algorithms diverge. Thus, a family of norm stochastic gradient adaptive filter algorithms was proposed, such as least mean absolute third (LMAT) algorithm \cite{zhao2014new,guan2017nonparametric}, least-mean-fourth (LMF) algorithm \cite{eweda2012stochastic,eweda2016stable}, and least-mean mixed-norm (LMMN) algorithm \cite{tanrikulu1996convergence,li2016sparse}. In \cite{chambers1997robust}, a robust mixed-norm (RMN) algorithm was developed based on a convex function of the error norms that underlie the least mean square (LMS) and least absolute difference (LAD) algorithms. Therefore, the RMN algorithm has robustness performance in the presence of impulsive noise.

To achieve improved performance in impulsive noise, several variants of the kernel adaptive filter were proposed \cite{miao2012kernel,wang2013kernel,wu2015kernel,zhao2017projected,gao2017kernel}. Particularly, in \cite{liu2014kernel}, the KRMN algorithm was proposed by deriving the RMN algorithm in RKHS. Unfortunately, the unsuitable selection of mixing parameter degrades the performance of KRMN algorithm. To overcome this problem, in this paper, we proposed two adaptation rules for the KRMN algorithm, called variable mixing parameter KRMN (VPKRMN). Based on the VPKRMN algorithms, we further proposed a quantized VPKRMN (QVPKRMN) algorithm to curb the growth of the networks. Furthermore, the energy conservation relation (ECR) and convergence property of QVPKRMN algorithm are analyzed. 

This paper is organized in the following manner. Section 2 introduces a brief description of the kernel method and the KRMN algorithm. In Section 3, two novel VPKRMN algorithms are proposed to adapt the mixing parameter. And, the QVPKRMN algorithm is proposed to control the growth of the kernel structure. In Section 4, the analysis of the convergence property is performed. Then, simulations in the context of nonlinear system identification are conducted in Section 5. Finally, conclusions are found in Section 6.

\section{Kernel method and KRMN algorithm}

\subsection{Kernel method}

The kernel method is a useful nonparametric modeling tools to deal with the nonlinearity problem. The power of this idea is to transform input data (input space $\mathcal{U}$) into a high-dimensional feature space $\mathcal{F}$ using a certain nonlinear mapping, which can be expressed as:
\begin{equation}
{\bm {\varphi}}:\;\; \mathcal{U} \to \mathcal{F}
\label{001}
\end{equation}
where $\bm {\varphi}$ is the feature vector in the kernel method. Based on Mercer theorem, a shift-invariant Mercer kernel can be expressed as \cite{liu2011kernel}:
\begin{equation}
\kappa ({\bm u},{\bm{u'}}) = {\bm{\varphi}}({\bm u}){\bm\varphi}^T({\bm{u'}}) = \sum\limits_{i = 1}^\infty {\phi_i\varphi_i({\bm u})\varphi_i^T({\bm{u'}})} 
\label{002}
\end{equation}
where $\phi$ is the nonnegative eigenvalue, and $\varphi$ is the corresponding eigenfunction. The eigenvalues and eigenfunctions constitute the feature vector $\bm{\varphi}$:
\begin{equation}
{\bm{\varphi}}({\bm u}) = {\left[{\sqrt {{\phi_1}} {\varphi_1}({\bm u}),\sqrt {{\phi_2}} {\varphi_2}({\bm u}),...} \right]^T}. 
\label{003}
\end{equation}

It is well known that a Mercer kernel is a continuous, symmetric and positive-definite kernel. The commonly used Gaussian kernel can be expressed as
\begin{equation}
\kappa ({\bm u},{\bm{u'}}) = \exp \left( { -h{{\left\| {{\bm u} - {\bm{u'}}} \right\|}^2}} \right)
\label{004}
\end{equation}
where $h$ is the kernel bandwidth. By using (\ref{002}) and (\ref{004}), the feature space can be calculated by inner product. Consequently, the output of adaptive filter can be expressed by inner product of the transformed test data $\varphi({\bm u})$ and training data $\varphi(\bm{u}_j)$ 
\begin{equation}
f({\bm u}) = \sum\limits_{j=1}^n {a_jy_j\langle \varphi ({\bm u}),} \varphi({\bm u}_j)\rangle 
\label{005}
\end{equation}
where $a_j$ is the coefficient at discrete time $n$, and $\langle \cdot \rangle$ is the inner product operation, respectively.

\subsection{KRMN algorithm}

When the desired or the input signal is corrupted by impulsive noise, the performance of the KLMS algorithm degrades. To overcome this problem, the KRMN algorithm was proposed by using the kernel method \cite{liu2014kernel}. The input data of RMN algorithm are transformed into RKHS as ${\bm\varphi}(n)$, and the weight vector in feature space is defined as ${\bf{\Omega}}(n)$, ${\bf{\Omega}}(1) = {\bf{0}}$. The error signal is defined as:
\begin{equation}
e(n) \triangleq d(n) - {{\bf{\Omega}}^T}(n){\bm\varphi}(n).
\label{006}
\end{equation}

The KRMN algorithm is based on minimization of the following mixed-norm error \cite{chambers1997robust,liu2014kernel}:
\begin{equation}
J(n) = \lambda E\left\{ e^2(n) \right\} + \left({1 - \lambda} \right)E\left\{{\left| e(n) \right|} \right\}
\label{007}
\end{equation}
where $E\left\{ \cdot \right\}$ is the statistical expectation operator, $\lambda$ is limited in the range of $(0,1)$. From (\ref{007}), the cost function of KRMN algorithm is a linear combination of the KLMS and KLAD \footnote{The KLMAD algorithm can be easily derived by casting the LAD algorithm into RKHS. When the mixing parameter of VPKRMN is equal to zero, the VPKRMN becomes KLAD.} algorithms. That is, the combination of $l_2$ norm and $l_1$ norm. The gradient vector of $J(n)$ with respect to ${\bf{\Omega}}(n)$ is 
\begin{equation}
{\nabla_{{\bf{\Omega}}(n)}}J(n) = -\left[ {2\lambda e(n) + (1 - \lambda)\mathrm {sign}\left\{ {e(n)} \right\}} \right]{\bm\varphi}(n)
\label{008}
\end{equation}
where $\mathrm {sign}\{x\}$ denotes the sign function, i.e., if $x > 0$, then $\mathrm {sign}\{x\}=1$, if $x=0$, $\mathrm {sign}\{x\}$ returns to 0, otherwise $\mathrm {sign}\{x\}=-1$. Hence, the adaptive rule of KRMN is solved iteratively on the new example sequence $\{ {\bm\varphi}(n),d(n)\}$
\begin{equation}
{\bf{\Omega}}(n + 1) = {\bf{\Omega}}(n) + \mu \left[ {2\lambda e(n) + (1 - \lambda)\mathrm {sign}\left\{ {e(n)} \right\}} \right]{\bm\varphi}(n).
\label{009}
\end{equation}
Reusing (\ref{009}), we have
\begin{equation}
{\bf{\Omega}}(n+1) = \mu \sum\limits_{j=1}^{n+1} {\left[ {2\lambda e(j) + (1 - \lambda)\mathrm {sign}\left\{ e(j) \right\}} \right]{\bm\varphi}(j)}.
\label{010}
\end{equation}
By using the Mercer kernel in (\ref{002}), the filter output can be calculated through kernel evaluations
\begin{equation}
y(n+1) = \mu \sum\limits_{j=1}^n {\left[ {2\lambda e(j) + (1 - \lambda)\mathrm{sign}\left\{ e(j) \right\}} \right]\kappa (j,n+1).} 
\label{011}
\end{equation}
For the sake of simplicity, we define
\begin{equation}
a_j(n+1) \triangleq \mu \left[ {2\lambda e(j) + (1 - \lambda)\mathrm{sign}\left\{ e(j) \right\}} \right],\;\;j = 1,...,n + 1
\label{012}
\end{equation}
and codebook ${\bf C}(n)$ refer as a center set in time $n$
\begin{equation}
{\bf C}(n+1) = \left[ {{\bf C}(n),{\bm u}(n+1)} \right].
\label{013}
\end{equation}
It can be observed that if the kernel function is replaced by a radial kernel, the KRMN algorithm produces a growing radial basis function (RBF) network by allocating a new kernel unit for every new example with input $\bm u(n+1)$. The main bottleneck of the KRMN algorithm is its network size grows with the number of processed data. To overcome this severe drawback, a quantization scheme should be used to curb the growth of network.

\section{Proposed algorithms}
\subsection{VPKRMN algorithm}

An unsuitable mixing parameter selection will lead to a performance degradation of the KRMN algorithm. To circumvent this problem, the mixing parameter $\lambda$ should be automatically adjusted. Here, we use $\lambda(n)$ instead of $\lambda$ to derive the variable mixing parameter algorithm. Considering $\lambda(n)$ to minimize the mixed-norm error of the KRMN algorithm at each iteration cycle, we obtain
\begin{equation}
\begin{aligned}
\lambda(n+1) =&\; \frac{{\partial J(n)}}{{\partial \lambda(n)}} \\ 
=&\; \frac{{\partial \left\{ {\lambda(n)E\left\{ {e^2(n)} \right\} + \left[ {1 - \lambda(n)} \right]E\left\{ {\left| {e(n)} \right|} \right\}} \right\}}}{{\partial \lambda(n)}} \\ 
\approx&\; \frac{{\partial \left\{ {\lambda(n)e^2(n) + \left[ {1 - \lambda(n)} \right]\left| e(n) \right|} \right\}}}{{\partial \lambda(n)}} \\ 
\end{aligned}
\label{014}
\end{equation}
where $\lambda(n)$ is restricted in $[0,1]$. Then, we add a scaling factor $\gamma$ to (\ref{014}) to control the steepness of $J(n)$. As a result, an adaptive update rules for KRMN algorithm is obtained, and we name the new algorithm the VPKRMN-Algorithm 1: 
\begin{equation}
\begin{aligned}
{\bf{VPKRMN - Algorithm\;1}}:\lambda(n + 1) =&\; \lambda (n) + \gamma \left\{ {E\left\{ {\left| {e(n)} \right|} \right\} - E\left\{ {e^2(n)} \right\}} \right\} \\ 
\approx&\; \lambda(n) + \gamma \left\{ {\left| e(n) \right| - e^2(n)} \right\}. \\ 
\end{aligned}
\label{015}
\end{equation}
From (\ref{015}), the mixing parameter is adjusted by switching the two types of error norm. When $|e(n)| > {e^2}(n)$, the mixing parameter tends to one, the KLMS algorithm plays a dominate role of the filter. When $|e(n)| < {e^2}(n)$, the mixing parameter tends to zero, the KLAD algorithm plays a dominate role of the filter. 
\begin{figure}[htb]
	\centering
	\includegraphics[scale=0.45] {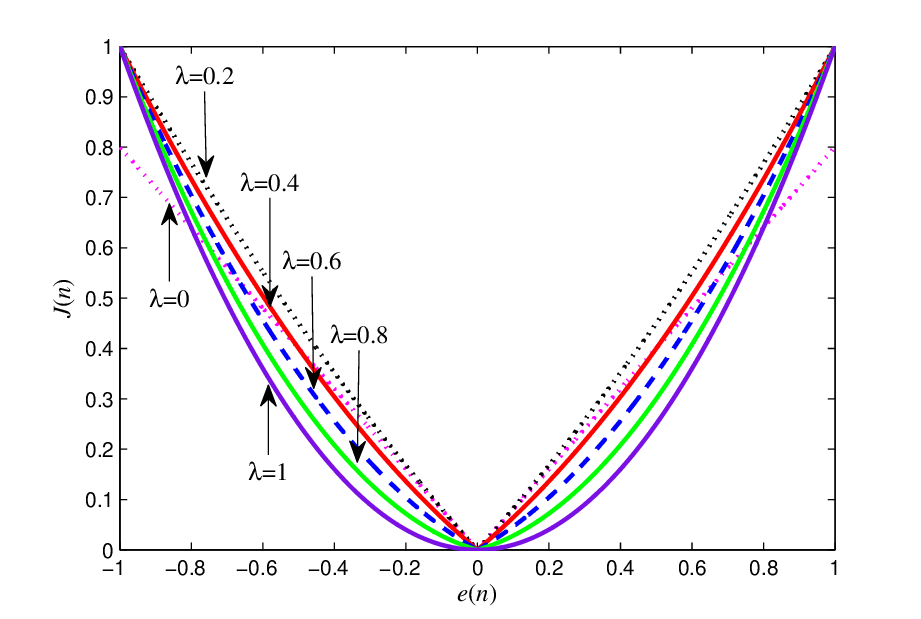}
	\caption{\label{1} The cost functions with different mixing parameter settings.}
	\label{Fig01}
\end{figure}

The cost function of the VPKRMN algorithm is a unimodal function (See Fig. \ref{Fig01}). The unimodal character is preserved for $\lambda(n)$ chosen in $(0,1)$, that is, the second term of (\ref{015}) keeps a very small value. Hence, the adaptation of VPKRMN Algorithm 1 is very sensitive to the choice of $\gamma$. To avoid this limitation, a new adaptive update approach, called VPKRMN-Algorithm 2, is proposed for adapt the mixing parameter of the KRMN algorithm.
\begin{equation}
{\bf{VPKRMN - Algorithm\;2}}:\left\{\begin{array}{l}
\lambda(n+1) = \delta \lambda(n) + \theta {p^2}(n) \\ 
p(n) = \beta p(n-1) + (1 - \beta)e(n)e(n-1) \\ 
\end{array} \right.
\label{016}
\end{equation}
where $\delta$ and $\beta$ are the exponential weighting parameters in the range of $[0,1]$, which control the quality of estimation of the algorithm, $\theta > 0$ is a positive constant, and $p(n)$ is a low-pass filtered estimation of $e(n)e(n-1)$. Note that the mixing parameter has a fixed value when $\delta=1$ and $\theta=0$. There are two reasons that account for the use of $p(n)$ in the update of $\lambda(n)$: (1) the error autocorrelation $e(n)e(n-1)$ is generally a good measure of the proximity to the optimum \cite{aboulnasr1997robust}; (2) the environment is divided into two cases by error autocorrelation $e(n)e(n-1)$: no impulsive environment and impulsive environment. The objective of VPKRMN-Algorithm 2 is to ensure the large $\lambda(n)$ when the VPKRMN Algorithm 2 is far from the optimum with decreasing $\lambda(n)$. The large value of $\lambda(n)$ leads to the $l_2$-norm error plays critical role, which provides a more accurate final solution and less misadjustment under no impulsive noise environment. Conversely, when algorithm suffers severely from the outlier problems, $\lambda(n)$ is small and the $l_1$-norm error offers a stable convergence characteristic to KRMN algorithm.

\subsection{QVPKRMN algorithm}

The QVPKRMN algorithm incorporates the idea of quantization into the VPKRMN algorithm to provide an efficient learning performance under impulse interference. In general, the quantization scheme is similar to the sparsification with NC method \cite{engel2004kernel}. In fact, they almost have the same computational complexity. The main difference between the quantization scheme and NC method is the quantization scheme utilizes the redundant data to locally update the coefficient of the closest center. The quantization method can be summarized as a learning strategy: the input space is quantized, if the current quantized input has already been assigned a center, no new center will be added, but the coefficient of that center will be updated through merging a new coefficient \cite{chen2012quantized}.

The feature vector ${\bm\varphi}(n)$ in quantization scheme can be expressed as:
\begin{equation}
\left\{\begin{array}{l}
{\bf{\Omega}}(0) = {\bf{0}} \\ 
e(n) = d(n) - {{\bf{\Omega}}^T}(n-1){\bm\varphi}(n) \\ 
{\bf{\Omega}}(n) = {\bf{\Omega}}(n-1) + \mu \left[ {2\lambda(n)e(n) + \left( {1 - \lambda(n)} \right)\mathrm {sign}\left\{ e(n) \right\}} \right]{\bm{Q}}\left[ {{\bm\varphi}(n)} \right] \\ 
\end{array} \right.
\label{017}
\end{equation}
where ${\bm Q}\left[\cdot\right]$ is a quantization operator in feature space $\mathcal{F}$. Owing to the high dimensionality of feature space, the quantization scheme is usually used in input space $\mathcal{U}$. Therefore, the learning rule of QVPKRMN algorithm in $\mathcal{U}$ can be given as:
\begin{equation}
\left\{ \begin{array}{l}
f_0 = 0 \\ 
e(n) = d(n) - f_{n-1}\left( {\bm u}(n) \right) \\ 
f_n = f_{n-1} + \mu \left[ {2\lambda(n)e(n) + \left( {1 - \lambda(n)} \right)\mathrm {sign}\left\{ {e(n)} \right\}} \right]\kappa \left( {Q[{\bm\varphi}(n)],\cdot} \right) \\ 
\end{array} \right.
\label{018}
\end{equation}
where $Q\left[\cdot\right]$ is a quantization operation in input space $\mathcal{U}$. Throughout this paper, the notation ${{\bm\varphi}_q}(n)$ is replaced by the notation $Q\left[ {{\bf{\varphi}}(n)} \right]$, ${{\bm u}_q}(n) = Q\left[{{\bm u}(n)}\right]$, ${{\bf{C}}_j}(n-1)$ is the $j$th element of ${{\bf{C}}}(n-1)$, $\left\| \cdot \right\|$ is the Euclidean norm in feature space $\mathcal{F}$, and ${\varepsilon_U}$ is the threshold of the distance. For $\varepsilon_U=0$, the QVPKRMN algorithm reduces to the VPKRMN algorithm. The proposed QVPKRMN algorithm is summarized in Table 1.

Table 2 summarizes the computational complexity of the algorithms, where $N$ is the training times, $M$ is the length of the filter, $L (L<N)$ is elements of index set. With an affordable computation complexity, the VPKRMN algorithm behaves much better than the KLMS and KLAD algorithms under the impulse noise environment. Since the QKLMS and QVPKRMN algorithms are developed by using quantization scheme, these algorithms have lower computation complexity than those of the KLMS, KLAD, KRMN and VPKRMN algorithms.

\begin{table}[tbp]
	\caption{Proposed QVPKRMN algorithms.}
	\small
	\centering
	\doublerulesep=0.4pt
	\begin{tabular}{lc}
		\hline
		\hline
		\textbf {Proposed QVPKRMN algorithms}\\
		\hline
		\hline
		\textbf{Initialization: choose step size} $\mu$\textbf{, bandwidth parameters of kernel} $h$\\
		${a_1}(1) = \mu {d^2}(1)sign\left\{ {d(1)} \right\}$,\;\;${\bf{C}}(1) = [{\bm{u}}(1)]$\\
		${f_1} = {a_1}(1)\kappa ({\bm{u}}(1), \cdot )$,\;\;$\lambda(1) = 0.5$\\
		\textbf{Computation:}\\
		While ${\bm u(n), d(n)}\;n>1$ is available do\\
		\;\;(1) Compute the output of the adaptive filter:
		\;\;$y(n) = \sum\limits_{j = 1}^{size({\bf{C}}(n-1))} {{{\bm{a}}_j}(n)\kappa ({\bm{u}}(n + 1),{\bm u}(j))}$\\
		\;\;(2) Compute the error: $e(n) = y(n)d(n)$\\
		\;\;(3) Compute the distance between $\bm u(n)$ and ${\bf{C}}(n-1)$\\		
\;\;\;\;\;\;$\mathrm {dis}({\bm u}(n),{\bf{C}}(n-1)) = \min\limits_{1 \le j \le size({\bf{C}}(n-1))} \left\| {{\bm{u}}(n) - {{\bf{C}}_j}(n-1)} \right\|$ \\
        \;\;(4) If $\mathrm {dis}\left( {{\bm u}(n),{\bf{C}}(n-1)} \right) \le {\varepsilon_U}$, keep the codebook unchanged:\\
\;\;\;\;\;\;${\bf{C}}(n+1) = {\bf{C}}(n)$\\
\;\;\;\;\;\;and quantize $\bm u(n)$ to the closest center through updating the coefficient of that center:\\      
\;\;\;\;\;\;${{\bf{a}}_{{j^*}}}(n) = {{\bf{a}}_{j^*}}(n+1) + \mu \left[ {2\lambda(n)e(n) + \left[ {1 - \lambda(n)} \right]\mathrm {sign}\left\{ e(n) \right\}} \right]$  \\
\;\;\;\;\;\;where ${j^*} = \min \limits_{1 \le j \le size({\bf{C}}(n-1))} \left\| {{\bm u}(n) - {{\bf{C}}_j}(n-1)} \right\|$ \\
\;\;\;\;\;\;otherwise, assign a new center and corresponding new coefficient:\\
\;\;\;\;\;\;${\bf{C}}(n+1) = [{\bf{C}}(n),{\bm u}(n+1)]$\\
\;\;\;\;\;\;${\bf{a}}(n) = \left[ {{\bf{a}}(n-1),\mu \left[ {2\lambda(n)e(n) + (1 - \lambda(n))\mathrm {sign}\{e(n)\}} \right]} \right]$ \\
Then, using two new update rule of mixing parameter\\
$\left\{ \begin{array}{l}
Algorithm\;1:\lambda(n+1) = \lambda(n) + \gamma \left\{ {\left| e(n) \right| - e^2(n)} \right\} \\ 
Algorithm\;2:\left\{\begin{array}{l}
\lambda(n+1) = \delta \lambda(n) + \theta p^2(n) \\ 
p(n) = \beta p(n-1) + (1 - \beta)e(n)e(n-1) \\ 
\end{array} \right. \\ 
\end{array} \right.$ \\
end while\\
		\hline
		\hline
	\end{tabular}
\end{table}

\begin{table}[tbp]
	\small
	\centering
	\caption{Summary of the computational complexity.}
	\doublerulesep=0.5pt
	\begin{tabular}{c|c|c|c|c}
		\hline
		\hline
		\textbf {Algorithms} &\textbf {Computation(training)} &\textbf {Memory(training)
		} &\textbf {Computation(test)} &\textbf {Memory(test)}\\ \hline 
		KLMS &$\mathcal{O}(N^2)$ &$\mathcal{O}(N)$ &$\mathcal{O}(N)$ &$\mathcal{O}(N)$\\ \hline
		KLAD &$\mathcal{O}(N^2)$ &$\mathcal{O}(N)$ &$\mathcal{O}(N)$ &$\mathcal{O}(N)$\\ \hline
		KRMN &$\mathcal{O}(N^2)$ &$\mathcal{O}(N)$ &$\mathcal{O}(N)$ &$\mathcal{O}(N)$\\ \hline
		QKLMS &$\mathcal{O}(L^2)$ &$\mathcal{O}(L)$ &$\mathcal{O}(L)$ &$\mathcal{O}(L)$\\ \hline
		\textbf{VPKRMN} &$\mathcal{O}(N^2)$ &$\mathcal{O}(N)$ &$\mathcal{O}(N)$ &$\mathcal{O}(N)$\\ \hline
		\textbf{QVPKRMN} &$\mathcal{O}(L^2)$ &$\mathcal{O}(L)$ &$\mathcal{O}(L)$ &$\mathcal{O}(L)$\\ \hline
	\end{tabular}
\end{table}

\section{Convergence analysis}

In this section, we establish the energy conservation relation (ECR) \cite{chen2012quantized,sayed2003fundamentals} for the QVPKRMN algorithm and analyze its mean convergence behavior. The convergence property of QVPKRMN algorithm is difficult to analyze exactly, so the theorem in \cite{mathews1987improved,vega2008new} and the independence assumption \cite{haykin2008adaptive} are introduced throughout the analyses. 

\subsection{Energy conservation relation}

Consider the adaptation of QVPKRMN algorithm in RKHS
\begin{equation}
{\bf{\Omega}}(n) = {\bf{\Omega}}(n-1) + \mu \left[ {2\lambda(n)e(n) + \left( {1 - \lambda(n)} \right)\mathrm {sign}\left\{e(n)\right\}} \right]{{\bm\varphi}_q}(n).
\label{019}
\end{equation}

We define the weight deviation vector ${\bm{V}}(n)$ and the second moment of the misalignment vector ${\bm{\eta}}(n)$ of the QVPKRMN algorithm as:
\begin{equation}
\begin{array}{l}
{\bm V}(n) \triangleq {\bf{\Omega}}(n) - {{\bf{\Omega}}_{opt}} \\ 
{\bm \eta}(n) \triangleq E\left\{ {{\bm V}(n){\bm V}^T(n)} \right\}. \\ 
\end{array}
\label{020}
\end{equation}
where ${{\bf{\Omega}}_{opt}}$ is the optimal weight vector. From (\ref{019}) and (\ref{020}), the update formulation of the weight deviation vector of QVPKRMN algorithm can be expressed as:
\begin{equation}
{\bm V}(n+1) = {\bm V}(n) - \mu \left[ {2\lambda(n)e(n) + \left[ {1 - \lambda(n)} \right]\mathrm {sign}\left\{ e(n) \right\}} \right]{{\bm\varphi}_q}(n).
\label{021}
\end{equation}

Then, we define the \emph{a posterior} error $e_p(n) \buildrel \Delta \over = {{\bm V}^T}(n){\bm{\varphi}}\left( {{\bm u}(n)} \right)$ and \emph{a priori error} $e_a(n) \buildrel \Delta \over = {{\bm V}^T}(n-1){\bm{\varphi}}\left( {{\bm u}(n)} \right)$. It can be shown that their \emph{a priori} and \emph{a posteriori} errors are related via
\begin{equation}
\begin{aligned}
e_p(n) =&\; e_a(n) - \mu \left[ {2\lambda(n)e(n) + \left[ {1 - \lambda(n)} \right]\mathrm {sign}\left\{ e(n) \right\}} \right]{{\bm{\varphi}}_q}(n) \\ 
=&\; e_a(n) - \mu \left[ {2\lambda(n)e(n) + \left[ {1 - \lambda(n)} \right]\mathrm {sign}\left\{ e(n) \right\}} \right]\kappa \left( {{{\bm u}_q}(n),{\bm u}(n)} \right). \\ 
\end{aligned}
\label{022}
\end{equation}

Combining (\ref{021}) and (\ref{022}) yields
\begin{equation}
{\bm V}(n) = {\bm V}(n-1) + \left[ {e_p(n) - e_a(n)} \right]\frac{{{{\bm{\varphi}}_q}(n)}}{{\kappa \left( {{{\bm u}_q}(n),{\bm u}(n)} \right)}}.
\label{023}
\end{equation}
Squaring both sides of (\ref{023}), we get
\begin{equation}
\begin{aligned}
{\bm V}^T(n){\bm V}(n) =&\; {\left[ {{\bm V}(n-1) + \left[ {e_p(n) - e_a(n)} \right]\frac{{{{\bm{\varphi}}_q}(n)}}{{\kappa \left( {{{\bm u}_q}(n),{\bm u}(n)} \right)}}} \right]^T} \\ 
&\cdot \left[ {{\bm V}(n-1) + \left[ {e_p(n) - e_a(n)} \right]\frac{{{{\bm{\varphi}}_q}(n)}}{{\kappa \left( {{{\bm u}_q}(n),{\bm u}(n)} \right)}}} \right]. \\ 
\end{aligned}
\label{024}
\end{equation}
Rearranging (\ref{024}), we have
\begin{equation}
\begin{aligned}
\left\| {\bm V}(n) \right\|_F^2 +&\; \frac{{e_a^2(n)}}{{{{\left[ {\kappa \left( {{{\bm u}_q}(n),{\bm u}(n)} \right)} \right]}^2}}} = \left\| {{\bm V}(n-1)} \right\|_F^2 \\ 
&+ \frac{{e_p^2(n)}}{{{{\left[ {\kappa \left({{{\bm u}_q}(n),{\bm u}(n)} \right)} \right]}^2}}} + {\beta_q} \\ 
\end{aligned}
\label{025}
\end{equation}
where ${\left\| \cdot \right\|_\mathcal{F}}$ is the norm in feature space $\mathcal{F}$, and $\beta_q = \frac{{2\left[ {e_p(n) - e_a(n)} \right]\left\{ {{\bm V}(n-1){{\bm{\varphi }}_q}(n)\kappa \left( {{{\bm u}_q}(n),{\bm u}(n)} \right) - e_a(n)} \right\}}}{{{{\left[ {\kappa \left( {{{\bm u}_q}(n),{\bm u}(n)} \right)} \right]}^2}}}$

As can be seen, (\ref{025}) of QVPKRMN algorithm is the same form as the QKLMS algorithm. When the quantization size goes to zero, $\beta_q \to 0$, the ECR expression for QKLMS algorithm is obtained 
\begin{equation}
\left\| {\bm V}(n) \right\|_\mathcal{F}^2 + e_a^2(n) = \left\| {\bm V}(n-1) \right\|_\mathcal{F}^2 + e_p^2(n).
\label{026}
\end{equation}

\subsection{Mean convergence}

In this subsection, the mean convergence analysis of weight vector is performed. Taking the mathematical expectation of (\ref{021}) and using independence assumption \cite{haykin2008adaptive}, we obtain
\begin{equation}
\begin{aligned}
E\left\{ {{\bm V}(n+1)} \right\} =&\; E\left\{ {{\bm V}(n)} \right\} - \mu E\left\{ {\left[ {2\lambda (n)e(n) + [1 - \lambda (n)]\mathrm {sign}\left\{ {e(n)} \right\}} \right]{{\bm{\varphi}}_q}(n)} \right\} \\ 
=&\; E\left\{ {{\bm V}(n)} \right\} - \left\{ {2\mu \lambda (n)E\left[ {e(n){{\bm \varphi}_q}(n)} \right] + \mu \left[ {1 - \lambda(n)} \right]E\left[ {\mathrm {sign}\left\{ e(n) \right\}{{\bm \varphi}_q}(n)} \right]} \right\}. \\ 
\end{aligned}
\label{027}
\end{equation}
According to \cite{mathews1987improved,vega2008new}, the second term of the right hand side in (\ref{027}) can be expressed as:
\begin{equation}
E\left[ {\mathrm {sign}\left\{ e(n) \right\}{{\bm{\varphi}}_q}(n)} \right] \approx \sqrt {\frac{2}{\pi}} \frac{1}{{{\sigma_e}}}E\left[ {e(n){{\bm \varphi}_q}(n)} \right].
\label{028}
\end{equation}

Substituting (\ref{028}) into (\ref{027}), we arrive
\begin{equation}
\begin{aligned}
E\left[ {{\bm V}(n + 1)} \right] \approx&\; E\left[ {{\bm V}(n)} \right] - \left\{ {2\mu \lambda (n)E\left[ {e(n){{\bm{\varphi}}_q}(n)} \right] + \mu \left[ {1 - \lambda(n)} \right]\sqrt {\frac{2}{\pi}} \frac{1}{{{\sigma_e}}}E\left[ {e(n){{\bm{\varphi}}_q}(n)} \right]} \right\} \\ 
\approx&\; E\left[ {{\bm V}(n)} \right]\left[ {1 - 2\mu \lambda(n) + \mu \left[ {1 - \lambda(n)} \right]\sqrt {\frac{2}{\pi }} \frac{1}{{{\sigma_e}}}} \right]E\left[ {{\bm{\varphi}}_q^T(n){{\bm{\varphi}}_q}(n)} \right]. \\ 
\end{aligned}
\label{029}
\end{equation}
where $e(n) \approx {\bm \varphi}_q^T(n){\bm V}(n)$. It is easily observed that $\bm V(n)$ will converge to zero vector as $n \to \infty$ if and only if the step size satisfies the following inequality
\begin{equation}
0 < \left\{ {2\mu \lambda(n) + \mu \left[ {1 - \lambda(n)} \right]\sqrt {\frac{2}{\pi}} \frac{1}{{{\sigma_e}}}} \right\}E\left[ {{\bm{\varphi}}_q^T(n){{\bm{\varphi}}_q}(n)} \right] < 2.
\label{030}
\end{equation}
Hence, we obtain
\begin{equation}
0 < \mu  < \frac{2}{{2\lambda(n) + \left[ {1 - \lambda(n)} \right]\sqrt {\frac{2}{\pi}} \frac{1}{{{\sigma_e}}}{{\bf{R}}_{{\bf{\varphi \varphi}}}}}}.
\label{031}
\end{equation}
where ${{\bf{R}}_{{\bm{\varphi \varphi}}}} = E\left[ {{\bm \varphi}_{\rm{q}}^T(n){{\bm \varphi}_q}(n)} \right]$. It is easy to see that the mean convergence condition of the QVPKRMN algorithm is 
\begin{equation}
0 < \mu  < \frac{2}{{2\lambda(n) + \left[ {1 - \lambda(n)} \right]\sqrt {\frac{2}{\pi}} \frac{1}{{{\sigma_e}}}{\lambda_{{\rm{max}}}}}}
\label{032}
\end{equation}
where $\lambda_{\max}$ is the maximum eigenvalues of ${{\bf{R}}_{{\bm{\varphi \varphi}}}}$. Since $\lambda_{\max} < \mathrm{tr}\left( {{{\bf{R}}_{{\bm{\varphi \varphi}}}}} \right)$, where $\mathrm{tr}\left( {{{\bf{R}}_{{\bm{\varphi \varphi}}}}} \right)$ denotes the trace of the autocorrelation matrix ${{\bf{R}}_{{\bm{\varphi \varphi}}}}$, a more rigorous condition can be gained 
\begin{equation}
0 < \mu  < \frac{2}{{2\lambda (n) + \left[ {1 - \lambda(n)} \right]\sqrt {\frac{2}{\pi}} \frac{1}{{\sqrt {{\zeta_{\min}}}}}\mathrm{tr}\left( {{{\bf{R}}_{{\bm{\varphi \varphi}}}}} \right)}}
\label{033}
\end{equation}
where $\zeta_{\min} = E\left\{ {{d^2}(n)} \right\} - {\bf{R}}_{{\bm{\varphi d}}}^T{{\bf{\Omega}}_{opt}}$, and ${\bf{R}}_{{\bm{\varphi d}}}$ is the cross-correlation vector of ${{\bm{\varphi}}_q}(n)$ and $d(n)$. The optimal weight vector can be expressed as:
\begin{equation}
{\bf{\Omega}}_{opt} = {\bf{R}}_{{\bm{\varphi \varphi }}}^{-1}{\bf{R}}_{{\bm{\varphi d}}}.
\label{034}
\end{equation}
From formula (\ref{019}), we get
\begin{equation}
\begin{aligned}
&E\left\{ {\left( {{{\bf{\Omega}}_{opt}} + {\bm V}(n+1)} \right){{\left( {{{\bf{\Omega}}_{opt}} + {\bm V}(n+1)} \right)}^T}} \right\} \\ 
=&\;E\left\{ {\left( {{{\bf{\Omega}}_{opt}} + {\bm{V}}(n)} \right){{\left( {{{\bf{\Omega}}_{opt}} + {\bm{V}}(n)} \right)}^T}} \right\} + \mu^2{{\bf{R}}_{{\bm{\varphi \varphi}}}} \\ 
&+ \mu E\left\{ {\left( {{{\bf{\Omega}}_{opt}} + {\bm{V}}(n)} \right){\bm \varphi}_q^T(n)K(n)} \right\} \\ 
&+ \mu E\left\{ {{{\bm{\varphi}}_q}(n){{\left( {{{\bf{\Omega}}_{opt}} + {\bm V}(n)} \right)}^T}K(n)} \right\} \\ 
\end{aligned}
\label{035}
\end{equation}
where $K(n) = 2\lambda(n)e(n) + \left[ {1 - \lambda (n)} \right]\mathrm {sign}\left\{ e(n) \right\}$. Thus, (\ref{035}) can be expressed with the form of the second moment of the misalignment vector
\begin{equation}
\begin{aligned}
{\bm{\eta}}(n+1) =&\; {\bm{\eta}}(n) + \mu^2{{\bf{R}}_{{\bm{\varphi \varphi}}}} + \mu E\left\{ {{\bm V}(n){\bm{\varphi}}_q^T(n)K(n)} \right\} \\ 
&+ \mu E\left\{ {{{\bm{\varphi}}_q}(n){\bm V}^T(n)K(n)} \right\}. \\ 
\end{aligned}
\label{036}
\end{equation}
Introducing (\ref{035}) to (\ref{036}) and using the independence assumption \cite{haykin2008adaptive}, (\ref{036}) can be given as:
\begin{equation}
\begin{aligned}
{\bm \eta}(n+1) =&\; {\bm \eta}(n) + \mu^2{{\bf{R}}_{{\bm{\varphi \varphi}}}} \\ 
&+ \mu E\left\{ {{\bm V}(n){\bm{\varphi}}_q^T(n)\left[ {2\lambda(n)e(n) + \left[ {1 - \lambda (n)} \right]\mathrm {sign}\left\{ e(n) \right\}} \right]} \right\} \\ 
&+ \mu E\left\{ {{{\bm \varphi}_q}(n){{\bm V}^T}(n)\left[ {2\lambda(n)e(n) + \left[{1 - \lambda (n)} \right]\mathrm {sign}\left\{ e(n) \right\}} \right]} \right\} \\ 
=&\; {\bm \eta}(n) + \mu^2{{\bf{R}}_{{\bm{\varphi \varphi}}}} + 2\lambda(n)\mu E\left\{ {{\bm V}(n){\bm{\varphi}}_q^T(n)e(n)} \right\} \\ 
&+ \left[ {1 - \lambda(n)} \right]\mu E\left\{ {{\bm V}(n){\bm{\varphi}}_q^T(n)\mathrm {sign}\left\{ e(n) \right\}} \right\} \\ 
&+ 2\lambda(n)\mu E\left\{ {{{\bm{\varphi}}_q}(n){\bm V}^T(n)e(n)} \right\} \\ 
&+ \left[ {1 - \lambda(n)} \right]\mu E\left\{ {{{\bm \varphi}_q}(n){\bm V}^T(n)\mathrm {sign}\left\{ e(n) \right\}} \right\}. \\ 
\end{aligned}
\label{037}
\end{equation}
Using the theorem in \cite{mathews1987improved,vega2008new}, the fourth term of equation (\ref{037}) can be respectively simplified as follows:
\begin{equation}
\begin{aligned}
&E\left\{ {{\bm V}(n){\bm{\varphi}}_q^T(n)\mathrm {sign}\left\{ e(n) \right\}} \right\} \\ 
=&\; E\left\{ {E\left\{ {{\bm V}(n){\bf{\varphi}}_q^T(n)\mathrm {sign}\left\{ e(n) \right\}|{\bm V}(n)} \right\}} \right\} \\ 
=&\; E\left\{ {{\bm V}(n)\sqrt {\frac{2}{\pi}} \frac{1}{{{\sigma_{e|{\bf{\Omega}}(n)}}}}E\left\{ {{\bm{\varphi}}_q^T(n)e(n)|{\bm V}(n)} \right\}} \right\} \\ 
=&\; E\left\{ {{\bm V}(n)\sqrt {\frac{2}{\pi }} \frac{1}{{{\sigma_{e|{\bf{\Omega}}(n)}}}}\left[ {{\bf{R}}_{{\bm{\varphi d}}}^T - {{\left[ {{{\bf{\Omega}}_{opt}} + {\bm V}(n)} \right]}^T}{{\bf{R}}_{{\bf{\varphi \varphi}}}}} \right]} \right\} \\ 
=&\; -E\left\{ {{\bm V}(n){{\bm V}^T}(n){{\bf{R}}_{{\bf{\varphi \varphi}}}}\sqrt {\frac{2}{\pi}} \frac{1}{{{\sigma_{e|{\bf{\Omega}}(n)}}}}} \right\} =  -\sqrt {\frac{2}{\pi}} \frac{1}{{{\sigma _e}}}{\bm{\eta}}(n){{\bf{R}}_{{\bm{\varphi \varphi}}}}. \\ 
\end{aligned}
\label{038}
\end{equation}
Similarity, the simplified form of sixth term of (\ref{037}) can be obtained
\begin{equation}
E\left\{ {{{\bm{\varphi}}_q}(n){\bm V}^T(n)\mathrm {sign}\left\{ e(n) \right\}} \right\} =  -\sqrt {\frac{2}{\pi}} \frac{1}{{{\sigma_e}}}{{\bf{R}}_{{\bm{\varphi \varphi}}}}{\bm \eta}(n).
\label{039}
\end{equation}
To calculate the third term and the fifth term of (\ref{037}), we have 
\begin{equation}
\begin{aligned}
E\left\{ {{{\bm V}^T}(n){{\bm{\varphi}}_q}(n)e(n)} \right\} =&\; E\left\{ {{{\bm{\varphi}}_q}(n){\bm V}^T(n)e(n)} \right\} \\ 
\approx&\; E\left\{ e^2(n) \right\} \buildrel \Delta \over = \sigma_e^2. \\ 
\end{aligned}
\label{040}
\end{equation}
Substituting (\ref{038}), (\ref{039}) and (\ref{040}) in (\ref{037}) will yield
\begin{equation}
\begin{aligned}
{\bm{\eta}}(n+1) =&\; {\bm \eta}(n) + \mu^2{{\bf{R}}_{{\bm{\varphi \varphi}}}} + 4\lambda(n)\mu \sigma_e^2 + \mu \left[ {1 - \lambda(n)} \right]\left[ { -\sqrt {\frac{2}{\pi}} \frac{1}{{{\sigma_e}}}{\bm{\eta}}(n){{\bf{R}}_{{\bm{\varphi \varphi}}}}} \right] \\ 
&+ \mu \left[ {1 - \lambda(n)} \right]\left[ { - \sqrt {\frac{2}{\pi}} \frac{1}{{{\sigma _e}}}{{\bf{R}}_{{\bf{\varphi \varphi}}}}{\bm \eta}(n)} \right] \\ 
=&\; {\bm \eta}(n)\left\{ {{\bf{I}} - \mu \left[ {1 - \lambda(n)} \right]\sqrt {\frac{2}{\pi }} \frac{1}{{{\sigma_e}}}{{\bf{R}}_{{\bm{\varphi \varphi}}}}} \right\} + {{\bf{R}}_{{\bm{\varphi \varphi}}}}\left\{ {\mu^2{\bf{I}} - \mu \left[ {1 - \lambda(n)} \right]\sqrt {\frac{2}{\pi}} \frac{1}{{{\sigma_e}}}{\bm \eta}(n)} \right\} \\ 
&+ 4\mu \lambda(n)\sigma_e^2{\bf{I}}. \\ 
\end{aligned}
\label{041}
\end{equation}

Furthermore, (\ref{041}) can be decomposed into a scalar form. The matrix $\bf M$ is defined as an orthonormal matrix of the autocorrelation matrix ${{\bf{R}}_{{\bm{\varphi \varphi}}}}$. Pre- and Post-multiplying both side of (\ref{041}) by $\bf M$ and ${\bf M}^T$, given
\begin{equation}
\begin{aligned}
{\bm \xi}(n+1) =&\; {\bm \xi}(n)\left\{ {{\bf{I}} - \mu \left[ {1 - \lambda(n)} \right]\sqrt {\frac{2}{\pi}} \frac{1}{{{\sigma_e}}}{\bm{\Lambda}}} \right\} \\ 
&+ {\bm{\Lambda}}\left\{ {{\mu^2}{\bf{I}} - \mu \left[ {1 - \lambda(n)} \right]\sqrt {\frac{2}{\pi }} \frac{1}{{{\sigma_e}}}{\bm{\xi}}(n)} \right\} + 4\mu \lambda(n)\sigma_e^2{\bf{I}}. \\ 
\end{aligned}
\label{042}
\end{equation}
where ${\bm \xi}(n)$ is a symmetric matrix, ${\bm \xi}(n) = {{\bf{M}}^T}(n){\bm \eta}(n){\bf{M}}(n)$, ${\bm{\Lambda}} = {{\bf{M}}^T}(n){{\bf{R}}_{{\bm{\varphi \varphi}}}}{\bf{M}}(n)$, and ${\bm{\Lambda}}$ is a diagonal matrix and its elements ${\lambda_i}(i - 1,2,...,M)$ are eigenvalues of matrix ${{\bf{R}}_{{\bm{\varphi \varphi}}}}$. A scalar form of (\ref{041}) can be obtained as:
\begin{equation}
\begin{aligned}
\xi_{ij}(n+1) =&\; \left\{ {1 - \mu \left[ {1 - \lambda(n)} \right]\sqrt {\frac{2}{\pi}} \frac{1}{{{\sigma_e}}}\left[ {{\lambda_i} + \lambda_j} \right]} \right\}\xi_{ij}(n) \\ 
&+ \mu^2\lambda_i\tau(i-j) + 4\mu \lambda(n)\sigma_e^2 \\ 
\end{aligned}
\label{043}
\end{equation}
where $\xi_{ij}(n)$ is the $(i,j)$th element of ${\bm \xi}(n)$, and $\tau(i-j) = \left\{ \begin{array}{l}
1,\;\;\;\;\;\mathrm {if}\;i = j \\ 
0,\;\;\;\;\mathrm {otherwise} \\ 
\end{array} \right..$ 

\section{Simulation results}
\begin{figure}[htb]
	\centering
	\includegraphics[scale=0.7] {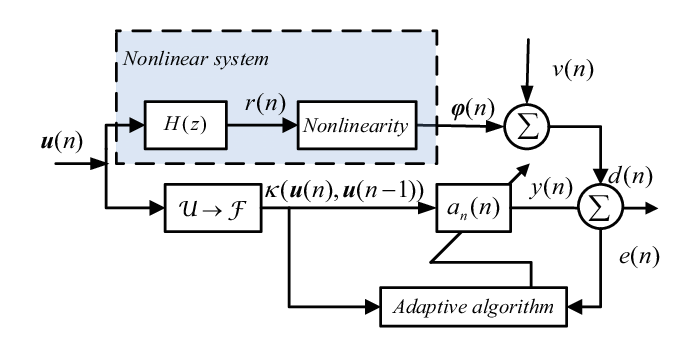}
	\caption{\label{2} Block diagram of the kernel adaptive identification.}
	\label{Fig02}
\end{figure}

To demonstrate the effectiveness of the proposed algorithms in impulsive noise environments, simulation studies are carried out for nonlinear system identification problem. In the following simulations, the software of Matlab R2013a is used to program the experiments under the computer environment of AMD (R) A10 CPU 2.1 GHz.

The block diagram of the kernel adaptive system identification is plotted in Fig. \ref{Fig02}. The goal of nonlinear system identification is to employ pairs of $\left\{ {{\bm{u}}(n),d(n)} \right\}$ inputs and addictive noise $v(n)$ to fit a function that maps an arbitrary system input into an appropriate output. The model coefficients at $n$ moment $a_n(n)$ are adjusted by the error signal  $e(n)$. The nonlinear system contains a linear filter and a memoryless nonlinearity. The linear system impulse response is generated by \cite{miao2012kernel}
\begin{equation}
\begin{aligned}
H(z) =&\; 0.1 + 0.2{z^{-1}} + 0.3{z^{-2}} + 0.4{z^{-3}} + 0.5{z^{-4}} \\ 
&+ 0.4{z^{-5}} + 0.3{z^{-6}} + 0.2{z^{-7}} + 0.1{z^{-8}} \\ 
\end{aligned}
\label{044}
\end{equation}
and the nonlinearity is given as $d(n) = r(n) - 0.9r^2(n) + v(n).$ 

\begin{figure}[htb]
	\centering
	\includegraphics[scale=0.5] {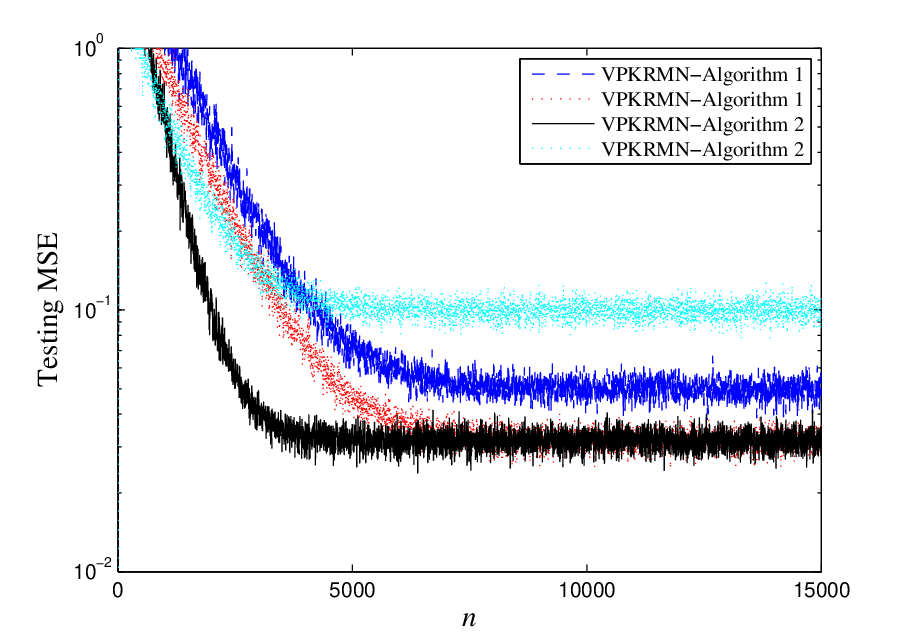}
	\caption{\label{3} The effect of the parameters on VPKRMN algorithms with $c=0.2$, $\sigma_I = \sigma_G=0.02$. (Keys: blue line, $\mu=0.1$, $\gamma=0.00001$; red line, $\mu=0.1$, $\gamma=0.00005$; black line, $\mu=0.1$, $\theta=0.01$, $\delta=0.97$, $\beta=0.98$; azury line, $\mu=0.1$, $\theta=0.05$, $\delta=0.97$, $\beta=0.98$.)}
	\label{Fig03}
\end{figure}

\begin{figure}[htb]
	\centering
	\includegraphics[scale=0.5] {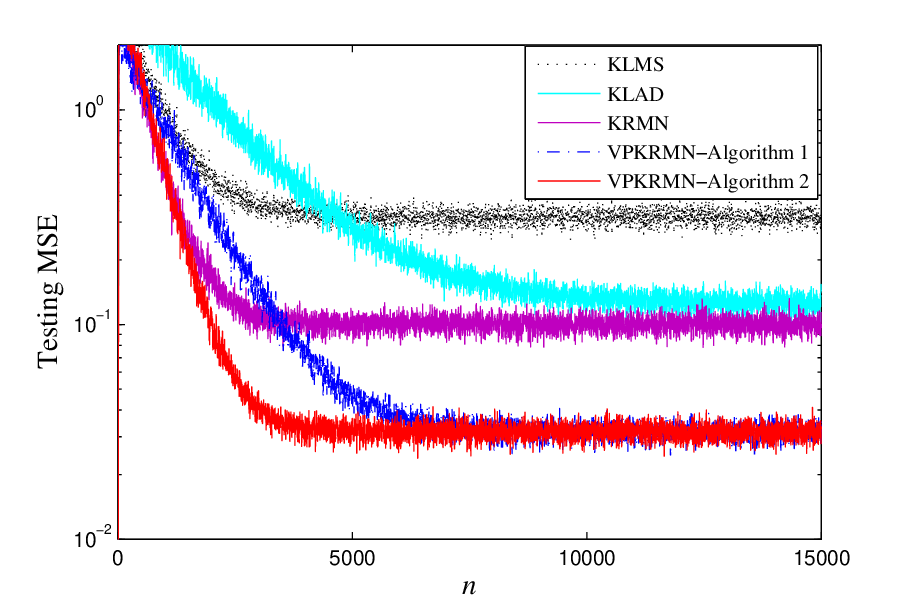}
	\caption{\label{4} Learning curves of KLMS, KLAD KRMN and VPKRMN algorithms for nonlinear system identification with $c=0.2$,  $\sigma_I = \sigma_G=0.02$. (Keys: KLMS, $\mu=0.1$, $h=0.05$; KLAD, $\mu=0.05$, $h=0.01$; KRMN, $\mu=0.1$, $h=0.1$, $\lambda=0.3$; VPKRMN-Algorithm 1, $\mu=0.1$, $h=0.05$;  VPKRMN-Algorithm 2, $\mu=0.1$, $h=0.1$.)}
	\label{Fig04}
\end{figure}

\begin{figure}[htb]
	\centering
	\includegraphics[scale=0.5] {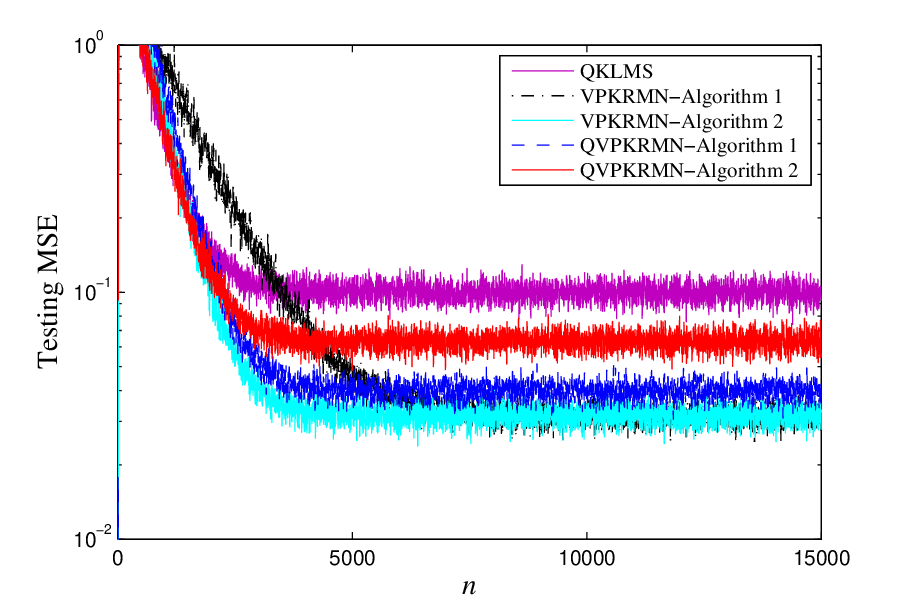}
	\caption{\label{5} Learning curves of QLMS, VPKRMN and QVPKRMN algorithms for nonlinear system identification with $c=0.2$,  $\sigma_I = \sigma_G=0.02$. (Keys: QKLMS, $\varepsilon_U=1$, $h=0.05$;  QVPKRMN-Algorithm 1, $\varepsilon_U=0.5$;  QVPKRMN-Algorithm 2, $\varepsilon_U=0.5$.)}
	\label{Fig05}
\end{figure}

\begin{figure}[htb]
	\centering
	\includegraphics[scale=0.5] {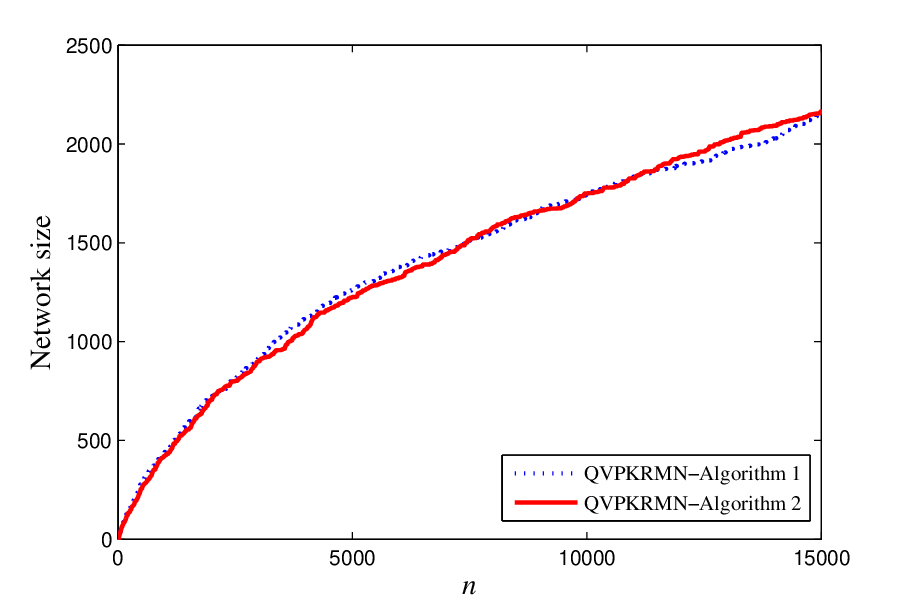}
	\caption{\label{6} Network size growth of the QVPKRMN algorithms.}
	\label{Fig06}
\end{figure}

\begin{figure}[htb]
	\centering
	\includegraphics[scale=0.5] {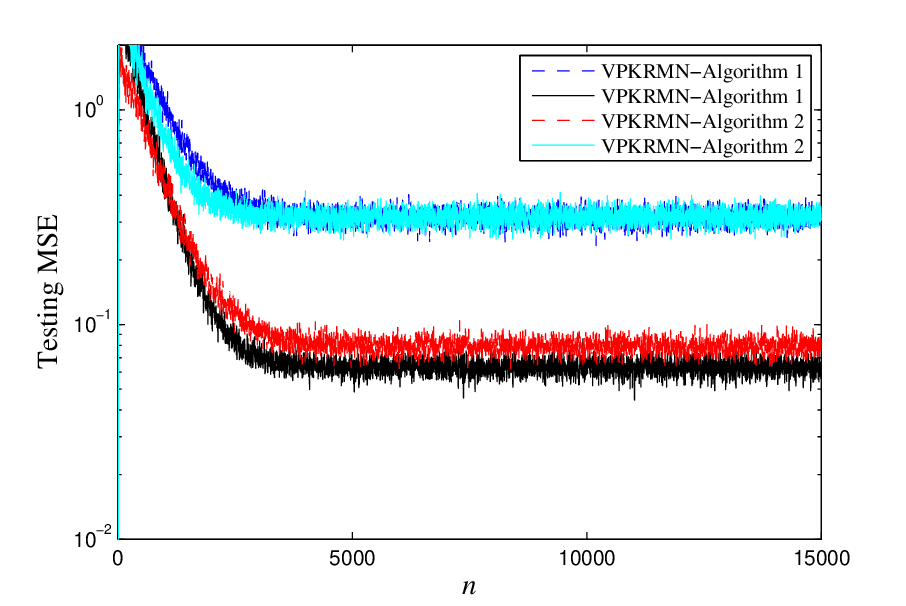}
	\caption{\label{7} The effect of the parameters for KRMN algorithms with $\alpha$-stable noise ($\alpha$=1.4, SNR=15dB). (Keys: blue line, $\mu=0.1$, $\gamma=0.00001$; black line, $\mu=0.1$, $\gamma=0.00003$; red line, $\mu=0.1$, $\theta=0.01$, $\delta=0.97$, $\beta=0.98$; azury line, $\mu=0.1$, $\theta=0.05$, $\delta=0.97$, $\beta=0.98$.)}
	\label{Fig07}
\end{figure}

\begin{figure}[htb]
	\centering
	\includegraphics[scale=0.5] {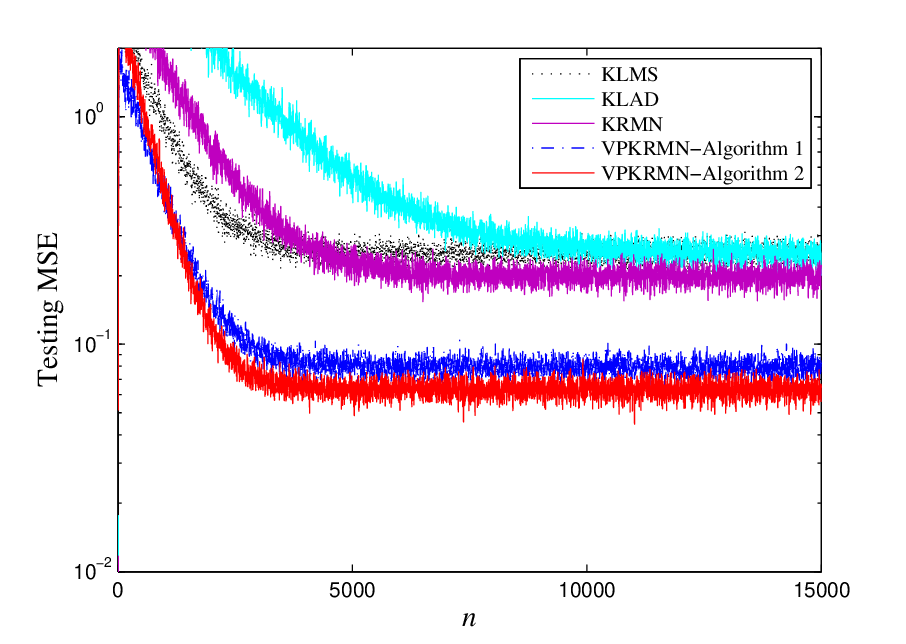}
	\caption{\label{8} Learning curves of KLMS, KLAD and VPKRMN algorithms for nonlinear system identification with $\alpha$-stable noise ($\alpha$=1.4, SNR=15dB). (Keys: KLMS, $\mu=0.1$, $h=0.05$; KLAD, $\mu=0.05$, $h=0.01$; KRMN, $\mu=0.1$, $h=0.1$, $\lambda=0.3$; VPKRMN-Algorithm 1, $\mu=0.1$, $h=0.1$;  VPKRMN-Algorithm 2, $\mu=0.1$, $h=0.1$.)}
	\label{Fig08}
\end{figure}

\begin{figure}[htb]
	\centering
	\includegraphics[scale=0.5] {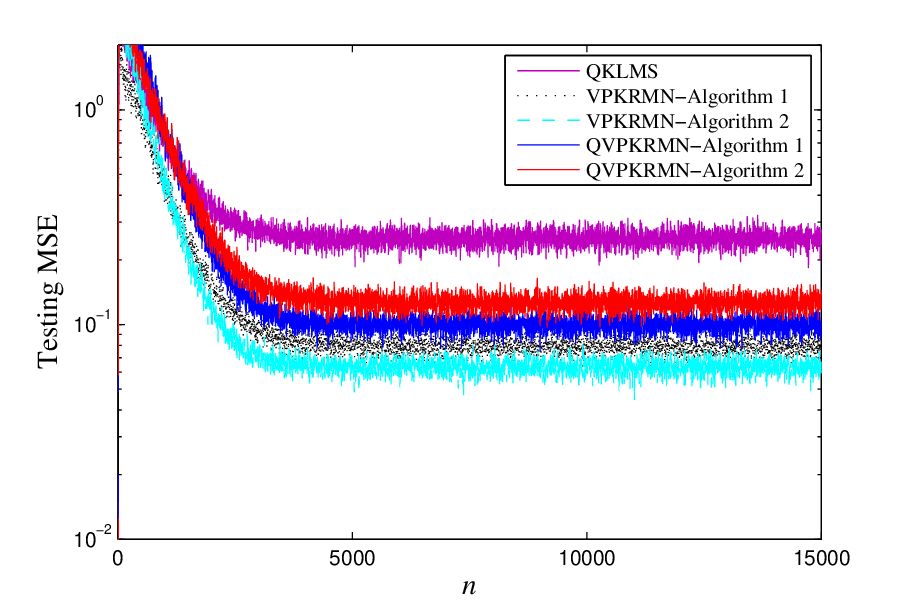}
	\caption{\label{9} Learning curves of QLMS, VPRMN and QVPKRMN algorithms for nonlinear system identification with $\alpha$-stable noise ($\alpha$=1.4, SNR=15dB). (Keys: QKLMS, $\varepsilon_U=1$, $h=0.1$;  QVPKRMN-Algorithm 1, $\varepsilon_U=0.01$;  QVPKRMN-Algorithm 2, $\varepsilon_U=0.1$.)}
	\label{Fig09}
\end{figure}

\begin{figure}[htb]
	\centering
	\includegraphics[scale=0.5] {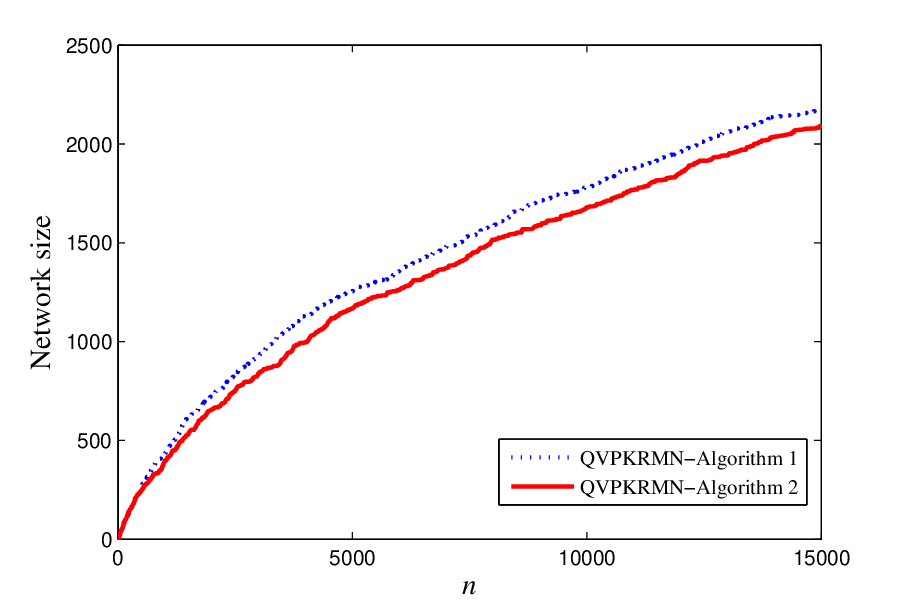}
	\caption{\label{10} Network size growth of the QVPKRMN algorithms.}
	\label{Fig10}
\end{figure}

\subsection{Test under impulsive noise environment with BG model}

In this example, the impulsive noise is modeled by Bernoulli-Gaussian (BG) distribution \cite{mathews1987improved} with probability function c and the root deviation $\sigma_I$. The white Gaussian noise (WGN) with zero mean and variance $\sigma_u^2=1$ is used as the input signal. The White Gaussian noise is a zero mean with root deviation $\sigma_G$. A segment of 15000 samples are used as the training data and another 1000 samples as the test data. Simulation results are obtained by 50 Monte Carlo trials. 

Firstly, the effect of the parameter on proposed VPKRMN algorithms are studied. Fig. \ref{Fig03} plots the effect of the update parameter for algorithm. It can be seen from this figure that the VPKRMN algorithm 1 achieves the fast convergence rate under $\gamma=0.00005$ as compared to $\gamma=0.00001$ of VPKRMN. And, the $\theta=0.01$ of KRMN algorithm 2 obtains the faster convergence speed than that of $\theta=0.05$. For this reason, the $\gamma=0.00005$ and $\theta=0.01$ are selected for proposed VPKRMN algorithm 1 and 2, respectively.

Figs. \ref{Fig04} and \ref{Fig05}illustrate the learning curves of existing algorithms. All the bandwidth parameters of kernel-based algorithms are set to 0.1. It observed from Fig. \ref{Fig04} that the proposed algorithms outperform the other algorithms in terms of convergence rate and steady-state error under the impulsive noise. The KRMN with $\lambda=0.3$ has fast convergence rate against impulse noise, but high MSE as compared to VPKRMN algorithms. Moreover, the VPKRMN-Algorithm2 achieves better performance than the VPKRMN-Algorithm1. Fig. \ref{Fig05} shows the performance of the proposed two algorithms based on quantization scheme, and the network size growth curves of QVPKRMN algorithms are plotted in Fig. \ref{Fig06}. As can be seen, the proposed QVPKRMN algorithms achieve faster convergence rate and lower MSE as compared to QKLMS algorithm, and slightly slow down the convergence rate as compared to the VPKRMN algorithms. Owing to using the quantization scheme, the QVPKRMN algorithms produce about 2000 network size in nonlinear system identification, which reduces the computational burden.

\subsection{Test under impulsive noise environment with $\alpha$-stable distribution model}

In the second example, the WGN is employed as the input signal, and the nonlinear system model in first experiment is continued to use. An impulsive noise can be modeled as the symmetric $\alpha$-stable distribution. A standard symmetric $\alpha$-stable ($S\alpha S$) distribution has the form \cite{shao1993signal}
\begin{equation}
{\varphi_{S\alpha S}}(t) = \exp \left\{ { -m{{\left| t \right|}^\alpha}} \right\}
\label{045}
\end{equation}
where $0<\alpha\leq2$ is a characteristic exponent, which indicates a peaky and heavy tailed distribution and likely more impulsive noise, and $m>0$ is dispersion of the noise. In following simulation studies, $\alpha=1.4$ is used, which is well model the radio frequency interference (RFI) for the embedded wireless data transceivers \cite{nassar2011mitigating}. 

In addition, the signal-to-noise ratio (SNR) of the $\alpha$-stable noise is defined as: \cite{zhao2015adaptive,lu2016collaborative}:
\begin{equation}
{\rm{SNR}} \triangleq \frac{\sigma_u^2}{m}.
\label{046}
\end{equation}

To demonstrate the effect of the variable parameter on the proposed algorithms, Fig. \ref{Fig07} shows the VPKRMN algorithms with different parameter settings. As can be found, a tiny change of the parameters cause a large change of the performance, and the appropriate selection of the parameters are $\gamma=0.0003$, $\theta=0.01$. Fig. \ref{Fig08} illustrates a comparison with the LMS, RMN, KLMS, KLAD, KRMN and VPKRMN algorithms for nonlinear system identification in $\alpha$-stable noise. Obviously, the KRMN algorithm has worse results than two VPKRMN algorithms because it is based on fixed mixing parameter, and the proposed VPKRMN algorithms achieve improved performance. Finally, we evaluate the performance of the QKLMS, VPKRMN and QVPKRMN algorithm, as shown in Fig. \ref{Fig09}. As can be seen, the proposed QVPKRMN algorithms have similar identification performance, and superior performance in the presence of $\alpha$-stable noise as compared to QKLMS algorithm. Fig. \ref{Fig10} shows the network size growth of QVPKRMN algorithms. One can see that the network size of QVPKRMN algorithm decreases to about 10$\%$ by sacrificing a little performance, which reduces the computational complexity.

From the experiment results of the above two examples, the proposed VPKRMN algorithms demonstrate the improved performance than the existing algorithms, and the performance of QVPKRMN algorithm is close to VPKRMN algorithm with less computational complexity. Also, the robustness of the proposed algorithms is confirmed by simulating various population sizes and different bandwidth parameters. The proposed VPKRMN algorithm 1 and VRKRMN algorithm 2 have similar misadjustment and convergence speed under the slightly impulsive process. By using the error autocorrelation $e(n)e(n-1)$, the VPKRMN algorithm 2 obtains a faster convergence rate than VPKRMN algorithm 1 in highly impulsive case. We conclude that all the proposed algorithms for nonlinear system identification can provide a satisfying result in impulsive interference.

\section{Conclusion}
Two VPKRMN algorithms and their quantization form (QVPKRMN algorithms) are proposed for nonlinear system identification under impulsive noises. The VPKRMN algorithms effectively solve the problem of mixing parameter selection. Then, to address the problem of computational intensive of the VPKRMN algorithm, the quantization scheme is introduced to the VPKRMN algorithms to generate a QVPKRMN algorithm. Moreover, the convergence behavior of the QVPKRMN algorithms is analyzed. Simulations results showed that the proposed VPKRMN algorithms are superior to the KLMS, KLAD and KRMN algorithms, and the QVPKRMN algorithm preserves the robustness performance under the impulsive interference with low computational complexity. 

\textbf{Acknowledgment}

The work partially supported by the National Science Foundation of P.R. China (Grant: 61571374, 61271340, 61433011). The first author would also like to acknowledge the China Scholarship Council (CSC) for providing him with financial support to study abroad (No. 201607000050).

\section*{References}

\bibliography{mybibfile}

\section*{Vitae}

\textbf{Lu Lu} is pursuing the Ph.D. degree in the field of signal and information processing at the School of Electrical Engineering, Southwest Jiaotong University, Chengdu, China. He is currently a Visiting Ph.D. Student with the Electrical and Computer Engineering, McGill University, Montreal, QC, Canada. His research interests include adaptive signal processing, kernel methods and evolutionary computing.
 
\textbf{Haiquan Zhao} was born in Henan, China, in 1974. He received the B.S. degree in applied mathematics, the M.S. degree, and the Ph.D. degree in signal and information processing from Southwest Jiaotong University, Chengdu, China, in 1998, 2005, and 2011, respectively. Since 2012, he has been a Professor with the School of Electrical Engineering, Southwest Jiaotong University. His current research interests include adaptive filtering algorithm, adaptive Volterra filter, nonlinear active noise control, nonlinear system identification, and chaotic signal processing.
 
\textbf{Badong Chen} received the B.S. and M.S. degrees in control theory and engineering from Chongqing University, in 1997 and 2003, respectively, and the Ph.D. degree in computer science and technology from Tsinghua University in 2008. He was a Post-Doctoral Researcher with Tsinghua University from 2008 to 2010, and a PostDoctoral Associate at the University of Florida Computational NeuroEngineering Laboratory (CNEL) during the period October 2010 to September 2012. He is currently a professor at the Institute of Artificial Intelligence and Robotics (IAIR), Xi'an Jiaotong University. His research interests are in signal processing, information theory, machine learning, and their applications in cognitive science and engineering. He has published 2 books, 3 chapters, and over 70 papers in various journals and conference proceedings. He is an IEEE senior member and an associate editor of IEEE Transactions on Neural Networks and Learning Systems and has been on the editorial boards of Applied Mathematics and Entropy

\end{document}